\documentclass[10pt,conference]{IEEEtran}
\IEEEoverridecommandlockouts

\usepackage{cite}
\usepackage{amsmath,amssymb,amsfonts}
\usepackage{algorithmic}
\usepackage{graphicx}
\usepackage{textcomp}
\usepackage{xcolor}

\def\BibTeX{{\rm B\kern-.05em{\sc i\kern-.025em b}\kern-.08em
    T\kern-.1667em\lower.7ex\hbox{E}\kern-.125emX}}
\begin{document}

%\title{Are Large Language Models really learning about Software Engineering?
%\title{Evaluating the Performance and Generality of LLMs in Software Engineering
\title{Augmenting the Generality and Performance of Large Language Models for Software Engineering
%\thanks{This work is funded by the German Research Foundation (DFG) through the project SENLP, grant 524228075.}
}

\author{\IEEEauthorblockN{Fabian C. Pe\~{n}a}
\IEEEauthorblockA{
\textit{Faculty of Computer Science and Mathematics} \\
\textit{University of Passau}\\
%Passau, Germany \\
fabiancamilo.penalozano@uni-passau.de}
}

\maketitle

\begin{abstract}
Large Language Models (LLMs) are revolutionizing software engineering (SE), with special emphasis on code generation and analysis. However, their applications to broader SE practices including conceptualization, design, and other non-code tasks, remain partially underexplored. This research aims to augment the generality and performance of LLMs for SE by (1) advancing the understanding of how LLMs with different characteristics perform on various non-code tasks, (2) evaluating them as sources of foundational knowledge in SE, and (3) effectively detecting hallucinations on SE statements. The expected contributions include a variety of LLMs trained and evaluated on domain-specific datasets, new benchmarks on foundational knowledge in SE, and methods for detecting hallucinations. Initial results in terms of performance improvements on various non-code tasks are promising.
\end{abstract}


\section{Introduction}

Research and applications in Natural Language Processing (NLP) have increased significantly with the arrival of the transformer architecture \cite{attention}, followed by some notable neural network models such as BERT \cite{bert}, GPT \cite{gpt3}, and T5 \cite{t5}. Nowadays, these and subsequent models are more commonly known as Large Language Models (LLMs). LLMs are designed to predict tokens (words, symbols, etc.) through a statistical process learned from data, emerging in higher-level capabilities or downstream tasks. In particular in the software engineering (SE) industry, LLMs have become popular, enabling tools such as GitHub Copilot \cite{copilot} and Devin \cite{devin} to assist developers in tasks related to code generation and analysis.

% scaling laws + coding capabilities + benchmarks

Newer LLMs tend to be larger and trained for longer on more extensive datasets, typically resulting in better performance \cite{scalinglaws, chinchilla}. For SE, learning from large amounts of source code, in different programming languages accompanied by natural language text with educational or industrial purposes, has emerged in capabilities such as writing code from free-text instructions or automatically generating technical documentation. Currently, these capabilities are partially evaluated using public benchmarks that compile thousands of coding problems with the corresponding unit tests for direct verification \cite{humaneval,mbpp,evalplus,swebench}.

% SE life cycle + surveys + opportunities

However, SE encompasses more than just writing or understanding code. The typical SE life cycle involves stages such as requirement analysis, software design, software development, quality assurance, and software maintenance. Consequently, researchers have also been concerned with building and evaluating LLMs across a broader range of SE tasks, as highlighted in recent surveys \cite{llm4se,llm4se2,llm4se3}. Despite the variety of datasets, network architectures, model sizes, prompting, and fine-tuning techniques documented in these surveys, there is a high imbalance toward coding-related tasks. Furthermore, there are no works focused on evaluating the foundational knowledge (i.e., factual and conceptual knowledge \cite{bloom}) of LLMs in SE.

% generality + AI systems + task-driven model choice

Building on the ideas presented in \cite{pathtoagi,comingwave} regarding the level of applicability that LLMs can potentially achieve, to effectively support SE practitioners in real-world scenarios, a certain level of generality (i.e., range of supported tasks) and performance are required. To satisfy this dual need, to the best of our knowledge, developers of LLM-based systems tend to use the most powerful general-purpose LLMs available on the market (e.g., GPT-4o, Claude 3.5, Llama 3.2), assuming them as the best choice for any type of task. We argue that this practice is far from optimal, since similar or better performance could potentially be achieved with smaller LLMs using different network architectures and trained on domain-specific datasets.

% hallucinations + opportunities

Another concern when working with LLMs is their risk of generating hallucinations (i.e., plausible but factually incorrect or nonsensical statements \cite{hallucinations}). Reducing this risk as much as possible is also essential to build trust among SE practitioners. While methods to detect and mitigate hallucinations have been proposed \cite{hallucinationssurvey1, hallucinationssurvey2}, none specifically target SE.

\section{Research questions}

On the path to generality and performance of LLMs for SE, with the assumption that achieving these properties should not be the goal of a single LLM from a system perspective, we identify opportunities to (1) advance the understanding of how LLMs with different pre-training datasets, network architectures, and model sizes perform on various non-code tasks, (2) build new benchmarks to evaluate LLMs as sources of foundational knowledge in SE, and (3) develop methods to effectively detect hallucinations on SE statements generated by LLMs. To address these opportunities, we propose the following research questions:

\textbf{Main research question:}
\begin{enumerate}
  \item{How much can the generality and performance of LLMs be augmented to cover a broader range of non-code and foundational knowledge SE-related tasks?}
\end{enumerate}

\textbf{Secondary research questions:}
\begin{enumerate}
  \setcounter{enumi}{1}
  \item{What is the impact of different domain-specific datasets, network architectures, and model sizes on the performance of LLMs for various non-code tasks identified in the literature?}
  \item {Under what scenarios can LLMs accurately provide foundational knowledge in SE in terms of terminology and common object-oriented design scenarios?}
  \item {Can effective methods be developed to detect hallucinations on SE statements generated by LLMs?}
\end{enumerate}

In recent years, LLM-based systems (a.k.a LLM agents) have gained important attention from the community as an alternative to everything that an LLM cannot achieve individually \cite{agents2}. LLM agents tend to outperform single LLMs on different tasks \cite{agents1,agents3} and, while the opportunities for these agents are extensive, fundamental capabilities of LLMs for SE at the most granular level remain partially underexplored. We argue that continuing to push for improvements in LLM agents with a strong focus only on coding-related tasks is potentially dangerous for new generations of SE practitioners and goes against what the software industry stands for.

\section{Expected contributions}

As part of this proposal, we expect to make four key contributions. Each contribution is accompanied by a peer-reviewed paper and a replication kit containing source code, datasets, and models. Network architecture, model size, training dataset, and refinement technique (i.e., prompting and fine-tuning, where applicable) are consistently considered as dimensions of analysis.

\subsection{LLMs trained on SE datasets}
\label{contrib1}

Based on common architectures like RoBERTa \cite{roberta}, GPT-2 \cite{gpt2}, and T5 \cite{t5}, among others, new LLMs are pre-trained from scratch and fine-tuned from publicly available checkpoints on over 200 GB of texts sourced from platforms such as GitHub, Stack Overflow, JIRA, and ArXiv. Subsequently, these LLMs are fine-tuned using a supervised approach on various non-code tasks identified in the literature. The performance on these tasks is evaluated using established metrics tailored to the specific task type. For benchmarking purposes, baseline models using machine learning algorithms like XGBoost and FastText are also trained. With these results we expect to contribute to the first and second research questions.

\subsection{Benchmarks on foundational knowledge in SE}
\label{contrib2}

To establish two new benchmarks on foundational knowledge in SE, terminology is extracted from several Standards Development Organizations (SDOs) \cite{standards, istqb, ireb, isaqb}, and a set of common object-oriented design scenarios with their corresponding UML diagrams are defined. LLMs trained in Section \ref{contrib1} are evaluated using these benchmarks in different settings. First, the ability to accurately discriminate between (i) terms and definitions, and (ii) design scenarios and UML diagrams. The main approach considered for the discrimination is zero-shot classification, which does not require specific fine-tuning of LLMs, and the evaluation is based on common classification metrics. Second, the performance in generating SE definitions and UML diagrams in formats like PlantUML\footnote{https://plantuml.com/}. Because LLMs may not be trained on the PlantUML syntax, fine-tuning or prompting techniques are optionally considered. The evaluation is based on metrics like Bilingual Evaluation Understudy (BLEU), additionally requiring a qualitative analysis to identify common reasons for errors. With these results we expect to contribute to the first and third research questions.

\subsection{Generation of nonsensical SE statements}
\label{contrib3}

Building upon the benchmarks established in Section \ref{contrib2}, a complementary benchmark of nonsensical SE statements is generated. Each entry consists of the original statement, a systematically modified nonsensical version, and the type of modification applied (e.g., replacement with antonyms, addition of an unsuitable negation, etc.). LLMs from Section \ref{contrib1} are prompted with these nonsensical statements to generate responses that may further propagate or elaborate on the nonsensical content. The evaluation primarily involves a qualitative assessment to determine if generation of nonsensical statements stems from deductive reasoning and if the underlying reasons align with the modifications applied to the prompts. The original statements are used as a control group. With these results we expect to contribute to the first and fourth research questions.

\subsection{Detection of nonsensical SE statements}

Utilizing the benchmark generated in Section \ref{contrib3}, an experimental process based on a zero-shot classification approach to detect nonsensical statements is conducted. This process involves classifying the original statements, the manually modified nonsensical statements, and the generated statements. The evaluation is carried out in multiple stages. First, for the original and modified statements, standard evaluation metrics are calculated. Second, for the generated statements, the label probabilities assigned by the LLM are compared to those assigned to the original and modified statements. The comparison is done by using techniques like QQ-plots and Kolmogorov-Smirnov test. Additionally, the analysis is complemented using a segmented approach based on the type of modification applied. With these results we expect to contribute to the first and fourth research questions.

\section{Initial results}

To date, BERT, RoBERTa, and GPT-2 models have been pre-trained from scratch and fine-tuned from publicly available checkpoints on more than 23 GB of textual data. Additionally, these models have also been fine-tuned on labeled datasets for 17 non-code tasks, showing an increase in performance in most cases. The first paper is being prepared for publication.

\section*{Acknowledgments}

This work is funded by the German Research Foundation (DFG) through the project SENLP, grant 524228075.

\bibliographystyle{IEEEtran}
\bibliography{refs}

\end{document}